\documentclass[prb,showpacs,twocolumn,aps,superscriptaddress,a4paper]{revtex4-1}
\usepackage{dcolumn,amssymb,amsmath,amsfonts,graphicx,latexsym,color}
\usepackage{epstopdf}
\begin{document}

\title{Phase diagram of the off-diagonal Aubry-Andr\'{e} model}

\author{Tong Liu}
\author{Pei Wang}
\author{Gao Xianlong}
\affiliation{Department of Physics, Zhejiang Normal University, Jinhua 321004, China}

\date{\today}

\begin{abstract}
We study a one-dimensional quasiperiodic system described by the off-diagonal
Aubry-Andr\'{e} model and investigate its phase diagram 
by using the symmetry and the multifractal analysis. It was shown in a recent work ({\it Phys. Rev. B}
{\bf 93}, 205441 (2016))
that its phase diagram was divided into three regions, dubbed the extended,
the topologically-nontrivial localized and the topologically-trivial localized phases, respectively.
Out of our expectation, we find an additional region of the extended phase which
can be mapped into the original one by a symmetry transformation.
More unexpectedly, in both ``localized'' phases,
most of the eigenfunctions are neither localized nor extended.
Instead, they display critical features, that is,
the minimum of the singularity spectrum is in a range $0<\gamma_{min}<1$ instead
of $0$ for the localized state or $1$ for the extended state.
Thus, a mixed phase is found with a mixture of localized and critical eigenfunctions.
\end{abstract}

\maketitle

\section{Introduction}
\label{n1}

Since Anderson's seminal paper in 1958~\cite{1an}
the metal-insulator transition has been studied in a wide range of
systems. The scaling theory~\cite{2scal} predicts
that there is no metal-insulator transition in one-dimensional (1D) systems with
randomly-distributed impurities.
On the other hand, 1D quasiperiodic systems~\cite{1PRB,2PRA,3PRA,He,Gramsch}
can host localized, extended or critical eigenstates.
The Aubry-Andr\'{e} (AA) model~\cite{7aubry} is an important paradigm of 1D quasiperiodic systems.
It can be derived from the reduction of a two-dimensional quantum Hall system
in the magnetic field~\cite{20SU}. Due to recent advances in experimental techniques, the AA model has been realized
in ultracold atoms~\cite{8BILLY,9ROATI} and photonic crystals~\cite{10PRL,11PRL}.
The phase diagram of the AA model
has been well understood with extensive researches~\cite{21SO,22SOU,23ZD,24GE,25MA,26WI}.
And many different variations of the AA model were studied also.
By including a long-range hopping term or modulating the on-site potentials,
some authors found a mobility edge in the spectrum which can
be precisely addressed by the duality symmetry~\cite{12PRL,13PRL}.
The others addressed a transition from the topological superconducting
phase to the localized phase in the AA model with p-wave pairing interaction~\cite{14PRL,15PRL,16PRB,Cao}.

Among different variations, the off-diagonal Aubry-Andr\'{e} model
displays an abundant phase diagram. It brings up either the zero-energy topological edge modes~\cite{17PRL}
or preserves the critical states in a large parameter space~\cite{18PRB},
depending on different ways of modulating the nearest-neighbor hopping amplitude.
\begin{figure}
	\centering
	\includegraphics[width=0.5
	\textwidth]{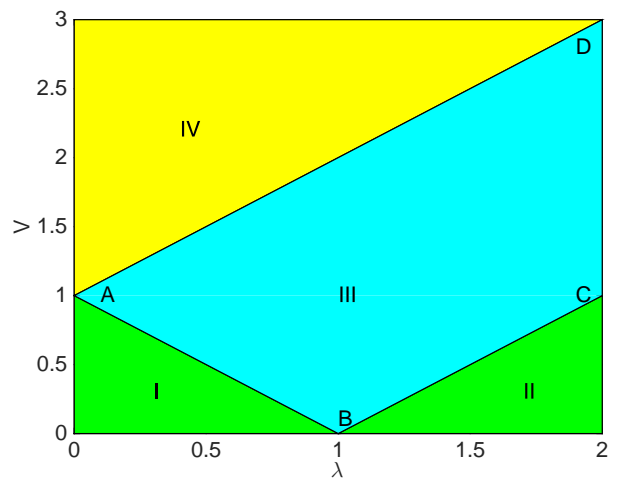}\\
	\caption{(Color online) Phase diagram of the off-diagonal Aubry-Andr\'{e}
	model. Four different regions are separated by the critical lines AB
	($V=1-\lambda$), BC ($V=\lambda-1$), and AD ($V=1+\lambda$).}
	\label{001}
\end{figure}
In a very recent paper~\cite{19PRB}, the authors
combined both the commensurate and incommensurate
modulations to explore the corresponding phase diagram. The model Hamiltonian
is expressed as
\begin{equation}\label{eq:ham}
    \hat H=-\sum_{i}^{L-1}(t+\lambda_{i}+V_{i})(\hat{c}_{i}^\dag \hat{c}_{i+1}+h.c.),
\end{equation}
where $\hat c_i$ is a fermionic annihilation operator,
$L$ is the total number of sites, and $\lambda_{i}=\lambda\cos(2\pi b{i})$
and $V_{i}=V\cos(2\pi\beta{i}+\phi)$ denote the commensurate and incommensurate
modulations, respectively. A typical choice of the parameters is $b=1/2$, $\beta=(\sqrt{5}-1)/2$
and $\phi = 0$. For convenience, $t = 1$ is set as the energy unit.

It was argued that the phase diagram of this model can be divided into
three regions, which are the extended, the topologically-nontrivial localized and the topologically-trivial localized phases, respectively.
In this paper we revisit this model by using the symmetry and the multifractal
analysis. Our findings are summarized in the phase diagram (Fig.~\ref{001}).
The main results
includes (i) there exists an additional region of extended phase (region~II)
which can be mapped into region~I by a newly-discovered symmetry transformation,
and (ii) region~III and~IV are mixed phases instead of localized phases
in which most of the eigenstates are critical states.

The rest of the paper is organized as follows. In Sec.~\ref{n2}, we
present the symmetry transformation for Hamiltonian~(\ref{eq:ham}), and
use it to determine the boundaries between different phases.
We further show that region~I and~II are in the extended phase
by calculating the inverse participation ratio numerically.
In Sec.~\ref{n3},
we apply the multifractal analysis in two different approaches.
In both approaches we verify that region~III and~IV are mixed phases with most of the eigenstates being critical states.

\section{Symmetry transformation and inverse participation ratio}
\label{n2}

We identify the phase boundaries of the off-diagonal Aubry-Andr\'{e} model
by finding its symmetry transformation.
For our purpose, the Schr\"{o}dinger equation is expressed every four sites as
\begin{widetext}
\begin{equation}
 \begin{split}
&  -\left(t+\lambda+V\cos\left(2\pi\beta\left(4n\right)\right)\right) \psi_{4n+1} - \left(
  t-\lambda +V\cos\left(2\pi\beta\left(4n-1\right)\right)\right) \psi_{4n-1} = E\psi_{4n} \\
&  -\left(t-\lambda+V\cos\left(2\pi\beta \left(4n+1\right)\right)\right) \psi_{4n+2} - \left(
  t+\lambda +V\cos\left(2\pi\beta\left(4n\right)\right)\right) \psi_{4n} = E\psi_{4n+1}\\
&  -\left(t+\lambda+V\cos\left(2\pi\beta \left(4n+2\right)\right)\right) \psi_{4n+3} - \left(
  t-\lambda +V\cos\left(2\pi\beta \left(4n+1\right)\right)\right) \psi_{4n+1} = E\psi_{4n+2} \\
&  -\left(t-\lambda+V\cos\left(2\pi\beta \left(4n+3\right)\right)\right) \psi_{4n+4} - \left(
  t+\lambda +V\cos\left(2\pi\beta \left(4n+2\right)\right)\right) \psi_{4n+2} = E\psi_{4n+3},
 \end{split}
\end{equation}
\end{widetext}
where $\psi_{j}$ denotes the eigenfunction in the first-quantization language and
$E$ is the eigenenergy. $n$ is an arbitrary integer.
We find a transformation with a period of four sites which keeps the Schr\"{o}dinger equation
invariant. The transformation is $t \to \lambda$, $\lambda \to t$, $\beta \to \beta +1/2$, $\psi_{4n}\to \psi_{4n}$,
$\psi_{4n+1}\to \psi_{4n+1}$, $\psi_{4n+2}\to -\psi_{4n+2}$ and $\psi_{4n+3}\to -\psi_{4n+3}$.
This transformation changes the sign of the wave function, but
has no influence on whether the wave function is localized or not. The shift of $\beta$ by $1/2$
keeps the absolute value of $V_j$ invariant at each site. Therefore, as we exchange $t$ and $\lambda$
while keeping $V$ invariant the wave function only changes the sign at some sites.
The transformation of $\left(t,\lambda,V\right)$
from $\left( 1,\lambda,V\right)$ to $\left( \lambda,1,V\right)$
does not change whether the eigenstate is localized or extended.
Furthermore, multiplying the Hamiltonian by an arbitrary number does not
change its eigenfunctions. As $t=1$ is fixed,
the simultaneous transformation $\lambda\to 1/\lambda$ and $V\to V/\lambda$
must relate two points in the phase diagram that belong to the same phase.

\begin{figure}
  \centering
  \includegraphics[width=0.5
  \textwidth]{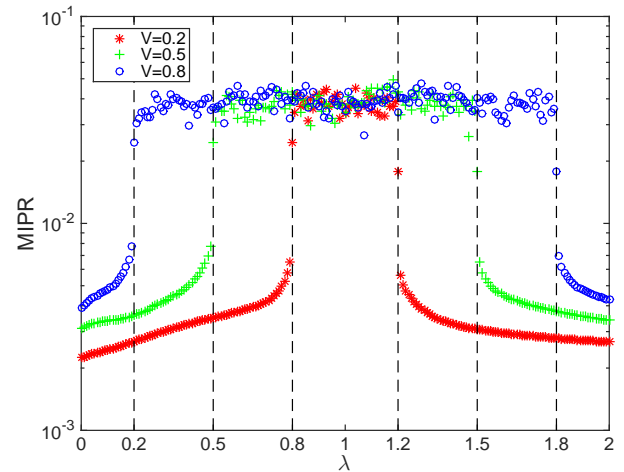}\\
  \caption{(Color online) MIPR as a function of $\lambda$ for different values of $V$.
  The dashed lines mark where MIPR changes abruptly. They are located at $\lambda = 1\pm V$.
  The total number of sites is set to $L=987$. We choose the open boundary condition.}
  \label{002}
\end{figure}
Under this transformation, region~I in the phase diagram (Fig.~\ref{001})
is mapped into region~II. For example, the boundary $AB$ ($V=-\lambda +1$), after the transformation, becomes $BC$ ($V=\lambda-1$). Especially, all the region below the boundary $AB$ can be mapped into the region below $BC$. Therefore, region~I and~II are dual and belong to the same phase.
The boundary $AD$ is $V=\lambda+1$. It keeps invariant under the symmetry transformation.
Region~III and ~IV are both mapped into themselves under the symmetry transformation.

By numerically calculating the inverse participation ratio (IPR) and the mean inverse participation ratio (MIPR), we find that region~I and~II are both the extended phase.
The IPR of a normalized wave function $\psi$ is defined as~\cite{27TH,28KO}
\begin{equation}
\text{IPR}_n =\sum_{j=1}^{L} \left|\psi^n_{j}\right|^{4},
\end{equation}
where $L$ denotes the total number of sites and $n$ is the energy level index.
It is well known that the IPR of an extended state scales like $L^{-1}$ which goes
to $0$ in thermodynamic limit. But for a
localized state IPR is finite even in thermodynamic limit.
For a critical state IPR scales as $L^{-\theta}$ with $0<\theta<1$.
There are $L$ different eigenfunctions for a specific Hamiltonian.
We then define the mean inverse participation
ratio (MIPR) as
\begin{equation}
\text{MIPR}=\sum_{n=1}^{L}\text{IPR}_{n}/L,
\end{equation}
where $\text{IPR}_{n}$ denotes the IPR of the $n$th eigenstate.

\begin{figure}
	\centering
	\includegraphics[width=0.5
	\textwidth]{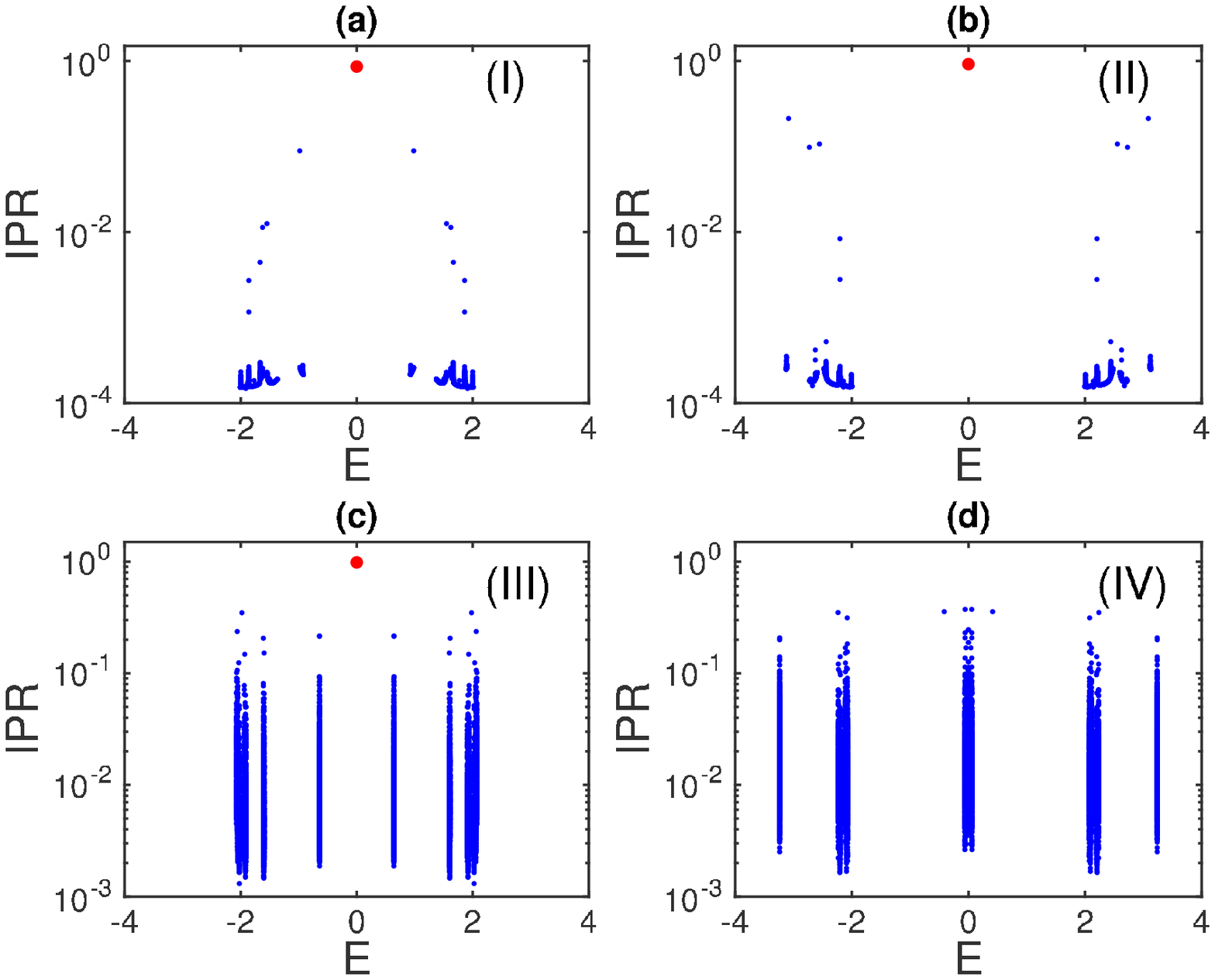}\\
	\caption{(Color online) The distribution of IPR among all the eigenstates for different
	$(\lambda,V)$ selected from the region (a) I with $(\lambda,V)=(0.5,0.2)$,
	(b) II with $(\lambda,V)=(1.5,0.2)$, (c) III with $(\lambda,V)=(0.5,0.55)$ and (d) IV with $(\lambda,V)=(0.5,1.55)$.
	The $x$-axis represents the eigenenergy $E$.
	The number of sites is set to $L=10000$.
	The red dots represent the zero-energy modes, which are present in
	the topologically-nontrivial phase.}
	\label{003}
\end{figure}
In Fig.~\ref{002}, we plot MIPR as a function of $\lambda$
at three disorder amplitudes: $V=0.2$, $V=0.5$, and $V=0.8$.
Here we used $L =987$. There are two turning points of MIPR
located at $\lambda = 1-V$ and $\lambda =1+V$, respectively.
At the turning points MIPR becomes very steep.
We also checked that with increasing number of sites $L$ the change of MIPR
at the turning points becomes even sharper. In the thermodynamic limit $L\rightarrow\infty$,
a singularity behavior of MIPR is expected signaling a transition between the extended phase and the localized (or critical) phase.
So the thermodynamically vanishing MIPR suggests that the system is in the extended regions for $\lambda < 1-V$ and $\lambda >1+V$.

We further study the distribution of IPR with different eigenstates.
The results are plotted in Fig.~\ref{003}.
We find the zero-energy modes (the red dots) in the region~I, II and~III.
Thus, these three regions are in the topologically-nontrivial phase.
But region~IV is topologically trivial.
Fig.~\ref{003}(a) and
Fig.~\ref{003}(b) plot the IPR distribution
in region~I and~II, respectively. The distribution has the same
characteristics in these two regions.
For almost all the eigenstates,
their IPR are close to each other, being very low (around $10^{-4}$).
This indicates that region~I and~II has a pure energy spectrum,
that all the eigenstates are extended. The extended, critical
and localized states do not coexist in these two regions.

On the other hand, region~III and~IV have the significantly
different IPR distribution (see Fig.~\ref{003}(c) and
Fig.~\ref{003}(d)). In these two regions, the value of IPR
is at least one order of magnitude larger than that of
the extended state ($> 10^{-3}$). At the same time,
the IPRs of different eigenstates disperse widely
in a range over two orders of magnitude (from $10^{-3}$ to
$10^{-1}$). Such a dispersed distribution suggests that
region~III and~IV should not be the pure localized phase.
Instead, there should exist critical states in these two regions.
To clarify the nature of them, we will apply the multifractal analysis
to the eigenfunctions in next section.

\section{Multifractal analysis}
\label{n3}

We carry out the multifractal analysis in two different approaches.
In the first approach, we fix the total number of sites in the system, while
dividing the system into a series of boxes with the box length tunable.
This is called the box-counting method~\cite{29SI,30HU}.
In the second approach, we
obtain the scaling behavior of the wave functions by changing
the total number of sites.

\subsection{Box-counting method}

\begin{figure}
  \centering
  \includegraphics[width=0.5\textwidth]{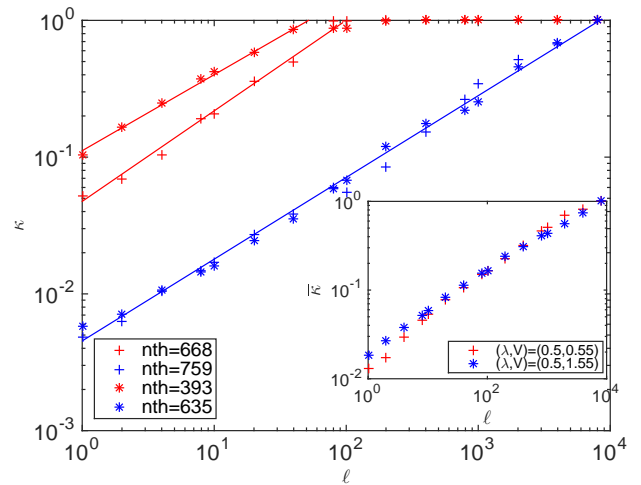}\\
  \caption{(Color online) $\kappa$ as a function of $l$ on the logarithmic scale.
  The crosses are for $(\lambda,V)=(0.5,0.55)$ which is located in region~III
  of the phase diagram. The stars are for $(\lambda,V)=(0.5,1.55)$ which is located in region~IV.
  Different colors represent different eigenstates.
  The inset plots $\overline{\kappa}$ which is the average of $\kappa$
  over all the eigenstates. The total number of sites is set to $L=8000$.}
  \label{004}
\end{figure}
Let us consider a normalized wave function $\psi$ defined over a chain of $L$ sites.
The probability of finding the particle at site $j$ is given by $p^n_j = \left|\psi^n_j\right|^2$
which satisfies $\sum_{j=1}^L p^n_j =1 $. In the multifractal analysis of wave functions,
$p^n_j\geq 0$ is viewed as an increase at site $j$ for the $nth$ eigenvalue. And the increase is supposed
to satisfy a local power law everywhere along the chain.

We divide the chain into $L/l$ segments with each segment containing $l$ sites.
The total increase in the $m$th segment is given by $P^n_m=\sum_{j=(m-1)l+1}^{ml} p^n_j$.
We then introduce a partition function
\begin{equation}
\kappa_n(q) = \sum_{m=1}^{L/l}\left( P^n_m\right)^q.
\end{equation}
The partition function obeys a power law $\kappa\sim (l/L)^{\tau}$ where the exponent $\tau$
is related to the multifractal dimension by $D(q)=\tau(q)/(q-1)$.
Following previous literatures we set $q=2$. It has been found that
$\tau(2)=D(2)$ tends to $0$ for a localized state but to $1$ for an extended state.
$0<\tau(2)<1$ signals a critical state.

We select two typical points in region~III and~IV, which are $\left(\lambda,V\right) = \left(0.5,0.55\right)$
and $\left(0.5,1.55\right)$, respectively.
Fig.~\ref{004} plots the corresponding $\kappa$ as a function of $l$ on the logarithmic scale.
By carefully checking the eigenstates in region III and IV, we find that the system has a mixed spectrum either of a localized or of a critical eigenstate. Here we take few typical eigenstates as examples.
For the $668$th eigenstate at $V=0.55$ (red crosses in Fig.~\ref{004}),
$\kappa$ scales as $l^{0.708}$ as $l<10^2$, but $\tau(2)$ tends to 0 for larger segment.
The existence of a turning point in the curve of $\kappa$ is the typical feature of a localized state.
Indeed, the localized states display multifractal feature up to the localization length.
Similarly, the $393$th eigenstate at $V=1.55$
is also a localized state (see red stars in Fig.~\ref{004}).
But for the $759$th eigenstate at $V=0.55$ (blue crosses) and the $635$th eigenstate
at $V=1.55$ (blue stars), $\kappa$ scales as $l^{0.613}$ over the whole domain of $l$.
The lack of turning point and a fractional $\tau$ together indicate that
these two states are critical. Therefore, in region~III and~IV the system
has a mixed spectrum in which the localized and critical states coexist.

We further study the average of $\kappa$ over different eigenstates, defined as
\begin{equation}
\overline{\kappa}=\sum_{n=1}^{L}\kappa_{n}(2)/L,
\end{equation}
where $\kappa_n$ denotes the partition function of the $n$th eigenstate.
The inset of Fig.~\ref{004} plots $\overline{\kappa}$ as a function
of $l$ on the logarithmic scale. In both region~III and~IV,
$\overline{\kappa}$ scales approximately as $l^\tau$ over the whole domain
of $l$. No obvious turning point is found. And $0<\tau<1$ is a fraction.
This indicates that most of the eigenstates in region~III and~IV
are critical states. Otherwise, $\overline{\kappa}$ would have a turning
point and display a platform for larger $l$. Because for larger $l$, $\tau(2)$
tends to 0 for a localized state but bigger than 0 for a critical state.
If there were a significant fraction of localized states in the spectrum,
their contribution to $\overline{\kappa}$ would dominate, and $\overline{\kappa}$
would then display a platform.

It is worth mentioning that the Legendre transformation of $\tau(q)$
is no more than the singularity spectrum. The latter is an important quantity
characterizing the multifractal nature of the system. We will discuss the
singularity spectrum in next subsection. Considering numerical stability,
we will employ a different method for calculating it.

\subsection{Finite size scaling}
\label{n4}

In the box-counting method, the segment length is tunable. An extreme case
is that each segment contains only a single site. The segment length is
then $1/L$. Note that the length of the whole chain is usually normalized to unity
in the multifractal analysis. Therefore, the segment length can also
be changed by changing the total number of sites $L$.
According to previous works~\cite{1PRB,31HI,32WA}, it is convenient to choose $L=F_m$
where $F_{m}$ is the $m$th Fibonacci number.
The advantage of this choice is that the golden ratio can be expressed as
$\beta = (\sqrt{5}-1)/2 =\lim_{m \rightarrow \infty} \frac{F_{m-1}}{F_{m}}$.

In a multifractal system the increase $p^n_j$ at any site satisfies a local power law:
\begin{equation}
p^n_j \sim \left(1/F_{m}\right)^{\gamma^n_j},
\end{equation}
where $\gamma^n_j$ is the singularity exponent. The set of sites
that share the same singularity exponent $\gamma$ is a fractal set of dimension $f(\gamma)$.
$f(\gamma)$ is just the so-called singularity spectrum. The approach of obtaining
$f(\gamma)$ is summarized as follows. The partition function is now defined as
$Z_m(q)=\sum_{j=1}^{F_m} \left(p^n_j\right)^q$. We then introduce the free energy
$G_m(q) = \ln Z_m(q)/m$. The singularity spectrum is the Legendre transformation
of the free energy, given by
\begin{equation}
 f(\gamma) = q\gamma - G_m(q)/\epsilon,
\end{equation}
with $\gamma=\displaystyle\frac{1}{\epsilon}\displaystyle\frac{d G_m}{dq}$ and $\epsilon=\ln\beta$.
For a critical wave function, $f(\gamma)$ is nonzero in an interval
$\left(\gamma_{min},\gamma_{max}\right)$ in which $f(\gamma)$ changes continuously.
In the thermodynamic limit $m\to \infty$, the value of $\gamma_{min}$ can be used
to distinguish the extended state ($\gamma_{min}=1$), the localized state ($\gamma_{min}=0$)
and the critical state ($0< \gamma_{min} <1$). Since there are $F_m$ eigenstates
for each Hamiltonian,
we use the average of $ \gamma_{min}$ over all the $F_m$ eigenstates as the indicator,
which is denoted by $\overline{\gamma_{min}}$.

\begin{figure}
  \centering
  \includegraphics[width=0.5\textwidth]{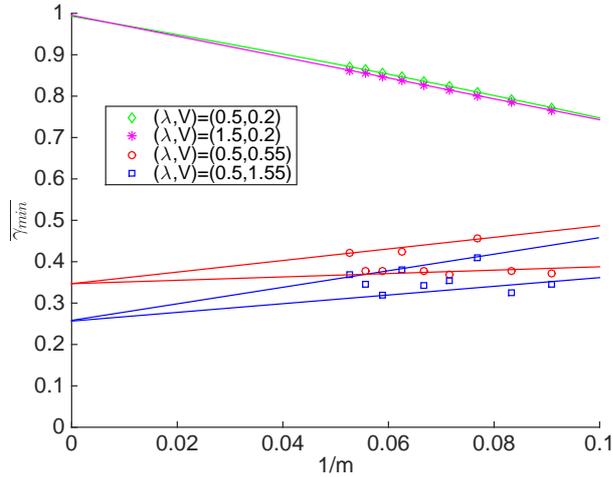}\\
  \caption{(Color online)  $\overline{\gamma_{min}}$ as a function of $1/m$ for
  $(\lambda,V)=(0.5,0.2)$, $(1.5,0.2)$, $(0.5,0.55)$ and $(0.5,1.55)$.
  These four points are located in region~I, II, III and IV, respectively.}
  \label{005}
\end{figure}

We plot $\overline{\gamma_{min}}$ as a function of $ 1/m $ for different $(\lambda,V)$
in Fig.~\ref{005}.
We find that $\overline{\gamma_{min}}$ extrapolates to $1$ for
$(\lambda,V)=(0.5,0.2)$ and $(1.5,0.2)$ which are both in the extended phase.
This fits our expectation.
$\overline{\gamma_{min}}$ extrapolates to $0.35$ for $(\lambda,V)=(0.5,0.55)$,
and to $0.27$ for $(\lambda,V)=(0.5,1.55)$.
The fractional $\overline{\gamma_{min}}$ is another evidence
that most of the states in region~III and~IV are critical states.

\section{Conclusions}
\label{n5}

In summary, we clarify the phase diagram of the off-diagonal
AA model. For this purpose, we discovered a symmetry transformation
which changes the sign of the wave function but keeps its amplitude invariant.
We also apply the box-counting and the finite size scaling methods
in analyzing the eigenfunctions. These two different approaches both
show that the wave functions display multifractal behavior.
We believe that the methods employed in this paper are useful
in a wide range of variations of the AA model.

\begin{acknowledgments}
This work was supported by the NSF of Zhejiang Province (Grant No. Z15A050001),
the NSF of China (Grant Nos. 11374266 and 11304280), and the Program for New Century Excellent Talents in University.

\end{acknowledgments}

\end{document}